\documentclass{article}

\usepackage{placeins}

\usepackage{subcaption}
\usepackage{arxiv}

\usepackage[utf8]{inputenc} % allow utf-8 input
\usepackage[T1]{fontenc}    % use 8-bit T1 fonts
\usepackage{hyperref}       % hyperlinks
\usepackage{url}            % simple URL typesetting
\usepackage{booktabs}       % professional-quality tables
\usepackage{amsfonts}       % blackboard math symbols
\usepackage{nicefrac}       % compact symbols for 1/2, etc.
\usepackage{microtype}      % microtypography
\usepackage{lipsum}		% Can be removed after putting your text content
\usepackage{graphicx}
\usepackage{doi}
\usepackage{CJKutf8}
\usepackage{multirow}

\title{Kiñit Classification in Ethiopian Chants, Azmaris and Modern Music: A New Dataset and CNN Benchmark}

\author{ \mbox{\href{https://orcid.org/0000-0003-0906-9605}{\includegraphics[scale=0.06]{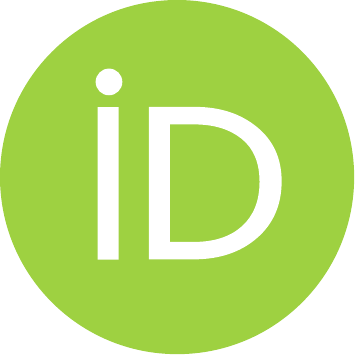}\strut}}\hspace{1mm}Ephrem Afele Retta \\
	School of Information Science and Technology\\
	Northwest University\\
	Xi’an 710127, China \\
	\texttt{afele@stumail.nwu.edu.cn} \\
	\And
	\mbox{\href{https://orcid.org/0000-0002-5549-5691}{\includegraphics[scale=0.06]{orcid.pdf}\strut}}\hspace{1mm}Richard Sutcliffe\thanks{Corresponding author} \\
	School of Information Science and Technology\\
	Northwest University\\
	Xi’an 710127, China \\
	School of Computer Science and Electronic Engineering \\ University of Essex \\
	Wivenhoe Park, Colchester CO4 3SQ, UK\\
	\texttt{rsutcl@nwu.edu.cn, rsutcl@essex.ac.uk} \\
	\And
	\mbox{\href{https://orcid.org/0000-0002-7182-9639}{\includegraphics[scale=0.06]{orcid.pdf}\strut}}\hspace{1mm}Eiad Almekhlafi \\
	School of Information Science and Technology\\
	Northwest University\\
	Xi’an 710127, China \\
	\texttt{ealmekhlafi@stumail.nwu.edu.cn} \\
	\And
	\mbox{\href{https://orcid.org/0000-0003-1393-2467}{\includegraphics[scale=0.06]{orcid.pdf}\strut}}\hspace{1mm}Yosef Kefyalew Enku \\
	School of Telecommunications Engineering\\
	Xidian University\\
	Xi’an 710071, China \\
	\texttt{enku@stu.xidian.edu.cn} \\
	\And
	\mbox{\href{https://orcid.org/0000-0002-9394-1613}{\includegraphics[scale=0.06]{orcid.pdf}\strut}}\hspace{1mm}Eyob Alemu \\
	School of Computer Science and Technology\\
	Xidian University\\
	Xi’an 710071, China \\
	\texttt{eyob@stu.xidian.edu.cn} \\
	\And
	\mbox{\href{https://orcid.org/0000-0003-3024-3620}{\includegraphics[scale=0.06]{orcid.pdf}\strut}}\hspace{1mm}Tigist Demssice Gemechu \\
	School of Information and Civil Engineering\\
	Chang'an University\\
	Xi’an 710064, China \\
	\texttt{2018128915@chd.edu.cn} \\
	\And
	\mbox{\href{https://orcid.org/0000-0002-5769-6040}{\includegraphics[scale=0.06]{orcid.pdf}\strut}}\hspace{1mm}Michael Abebe Berwo \\
	School of Information Science and Technology\\
	Chang'an University\\
    Xi’an 710064, China \\
	\texttt{2019024902@chd.edu.cn} \\
	\And
	\mbox{\href{https://orcid.org/0000-0002-3106-669X}{\includegraphics[scale=0.06]{orcid.pdf}\strut}}\hspace{1mm}Mustafa Mhamed \\
	School of Information Science and Technology\\
	Northwest University\\
	Xi’an 710127, China \\
	\texttt{mustafamhamed@stumail.nwu.edu.cn} \\
	\And
	\mbox{\href{https://orcid.org/0000-0002-0706-2103}{\includegraphics[scale=0.06]{orcid.pdf}\strut}}\hspace{1mm}Jun Feng\thanks{Corresponding author}\\
	School of Information Science and Technology \\
	Northwest University\\
	Xi’an 710127, China \\
	\texttt{fengjun@nwu.edu.cn} \\
}

\date{}

\renewcommand{\headeright}{E. A. Retta et al.}
\renewcommand{\undertitle}{}
\renewcommand{\shorttitle}{Kiñit Classification}

\hypersetup{
pdftitle={A template for the arxiv style},
pdfsubject={q-bio.NC, q-bio.QM},
pdfauthor={David S.~Hippocampus, Elias D.~Striatum},
pdfkeywords={First keyword, Second keyword, More},
}

\begin{document}
\maketitle

\begin{abstract}
	%\lipsum[1]
In this paper, we create EMIR, the first-ever Music Information Retrieval dataset for Ethiopian music. EMIR is freely available for research purposes and contains 600 sample recordings of Orthodox Tewahedo chants, traditional Azmari songs and contemporary Ethiopian secular music.
Each sample is classified by five expert judges into one of four well-known Ethiopian Kiñits, Tizita, Bati, Ambassel and Anchihoye. Each Kiñit uses its own pentatonic scale and also has its own stylistic characteristics. Thus, Kiñit classification needs to combine scale identification with genre recognition.
After describing the dataset, we present the Ethio Kiñits Model (EKM), based on VGG, for classifying the EMIR clips.
In Experiment 1, we investigated whether Filterbank, Mel-spectrogram, Chroma, or Mel-frequency Cepstral coefficient (MFCC) features work best for Kiñit classification using EKM. MFCC was found to be superior and was therefore adopted for Experiment 2, where the performance of EKM models using MFCC was compared using three different audio sample lengths. 3s length gave the best results. In Experiment 3, EKM and four existing models were compared on the EMIR dataset: AlexNet, ResNet50, VGG16 and LSTM. EKM was found to have the best accuracy (95.00\%) as well as the fastest training time. We hope this work will encourage others to explore Ethiopian music and to experiment with other models for Kiñit classification.

\end{abstract}

% keywords can be removed

\section{Introduction}
%\lipsum[2]
%\lipsum[3]
Music is an important part of everyday life. Around the world, it exists in many different forms and styles. Because musical preferences vary from person to person, categorizing music and making recommendations to listeners has become an important research topic \cite{elbir2020music} with many applications in listening apps and other platforms \cite{elbir2018music}.
\emergencystretch=3em
Multimedia file production and sharing through different mediums is increasing enormously.
In consequence, indexing, browsing, and retrieval of music files has become challenging and time-consuming. Numerous digital music classification techniques have been introduced \cite{doraisamy2008study}, but the majority of them are only developed and tested on well-known Western music datasets.
In Ethiopia, music classification is still being performed by individual music experts for archival or related purposes. Because of the amount of Ethiopian music now available in digital form, classification cannot be carried out with sufficient speed. As a result, even though the composer Saint Yared flourished in Ethiopia during the 6th Century \cite{shelemay1993oral} (p71), some five hundred years before Hildegard of Bingen \cite{white1998musical} the music of this country is not well known elsewhere.
In Ethiopia, music is based around several types of scale. Among these, four pentatonic scales (Kiñits) are particularly important \cite{abate2009ethiopian}\cite{assefa2009significance}: Anchihoye, Ambassel, Bati and Tizita. Because the music written in each Kiñit has its own characteristic style and features, the task of Kiñit classification is closely related to that of genre classification in European music.
A major challenge for Ethiopian Kiñit classification is the absence of training data. We have addressed this by creating the Ethiopian Music Information Retrieval (EMIR) dataset which includes data for the four main Kiñits.
We have also developed the Ethio Kiñits Model (EKM), a genre classification model based on the well-known VGG architecture. We then carried out three experiments. The first experiment selected an appropriate method from the FilterBank, Mel-spectrogram (MelSpec), Chroma, and Mel-frequency Cepstral Coefficient (MFCC) technologies for extracting features from recordings in our EMIR dataset.  MFCC was found to be the most effective in terms of accuracy and training time. The second experiment measured the effectiveness of different sample lengths for genre classification, in order to find the optimal length. The third experiment compared the classification performance of EKM and four other popular models using  MFCC features, working with EMIR datasets.

The contributions of this paper are as follows:
\begin{itemize}
\item We create for the very first time a dataset for Ethiopian music scales (Kiñits), called EMIR. There are 600 music samples, 162 Tizita, 144 Bati, 147 Ambassel, and 147 Anchihoye.
\item Five judges evaluate the recordings, and agreement between them is high (Fleiss kappa = 0.85). So the data is of high quality.
\item We develop a high-performing variant of the VGG MIR model which has just four CNN layers. We call this EKM.
\item We compare Filterbank, MelSpec, Chroma and Mel-frequency Cepstral Coefficient (MFCC) features and show experimentally in an MIR task that  MFCC leads to higher accuracy, using the proposed EKM model and our EMIR data. 
\item We compare the performance of EKM using different audio sample lengths, namely one second, three seconds and five seconds, working with EMIR data. Three seconds results in the best performance.
\item We apply EKM and four other architectural models to the MIR task, working with EMIR, and show that EKM is very effective, and by far the fastest.

\end{itemize}

The rest of this paper is organized as follows: Section 2 reviews previous work on music genre classification. Section 3 presents the EMIR dataset, describing the rationale behind its design and the method by which it was created. Section 4 discusses feature extraction for MIR, briefly outlining Filterbank, MelSpec, Chroma and MFCC. Section 5 describes the methodology, EKM architecture and settings used for our experiments. Section 6, presents the experiments and results. Finally, Section 7 gives conclusions and next steps. 

\section{Previous work on Music Genre Classification}
According to Tzanetakis and Cook \cite{tzanetakis2002musical}
in their landmark article, genres are categorial labels which classify pieces of music, based on instrumentation, rhythmic structure and harmonic content. The authors deduced three essential features for musical content, namely timbral texture, rhythm, and pitch content for Western music in various styles, including classical, jazz, pop and rock. This work paved the way for further research in the area of genre classification. Either whole recordings or homogeneous sections within them were used, and a classification accuracy of 61\% was achieved for ten genres, using statistical pattern recognition classifiers. The results closely matched those reported for human genre classification.

Jothilakshmi \cite{jothilakshmi2012automatic} applied a Gaussian mixture model (GMM), and a K-Nearest Neighbor (KNN) algorithm with spectral shape and perceptual features to Indian music datasets containing five genres. GMM gave the best accuracy with 91.25\%.
Rajesh \cite{rajesh2016automatic} again utilized KNN and support vector machines (SVM), using different feature combinations. The highest classification recorded (96.05\%) was by SVM using fractional MFCC with the addition of spectral roll off, flux, skewness, and kurtosis.

Al Mamun \cite{al2019bangla} used both deep learning and machine learning approaches on Bangla music datasets with six genres. The neural network model performed best compared to the machine learning methods, with accuracy 74\%.
Folorunso \cite{folorunso2021dissecting} investigated KNN, SVM, eXtreme Gradient Boosting (XGBoost) and Random Forest on Nigerian songs (ORIN dataset) with five genres. The XGBoost classifier had the highest accuracy (81.94\%).
De Sousa \cite{de2016robust} implemented SVMs on a Brazilian Music Dataset (BMD) with seven genres. The set of features they proposed was specifically tailored to genre recognition yielding a high classification accuracy of 86.11\%.

Kızrak \cite{saugun2016classification} used Deep Belief Networks (DBNs) to classify the music genre of Turkish classical music Makams, % CTM == (Classical Turkish Music) 
working with seven Makam datasets. Mel Frequency Cepstral Coefficients (MFCC) were employed on the collection of features, resulting in a classification accuracy of 93.10\%.
Thomas and Alexander \cite{lidy2016parallel} used a parallel Convolutional Neural Network (CNN) to identify the mood and genre of a song. They employed Mel-Spectograms which were extracted from audio recordings, and applied a CNN to accomplish their desired task.
Ali and Siddiqui \cite{ali2017automatic} implemented a machine-learning algorithm to classify music genre, using KNN and SVM. To obtain information from individual songs they extracted MFCCs from audio files.
Panteli et al. \cite{panteli2017computational} use MFCC features and traditional machine learning to analyse recordings of world music from many countries with the aim of identifying those which are distinct.
Phan et al. \cite{phan2021multi} carry out music classification in terms of environmental sound, audio scene and genre. They use four CRNN models, using Mel, Gammatone, CQT and Raw inputs. The outputs are combined to produce the classification.
Ma et al. \cite{ma2021computational} aim to predict the genre of a film using Music Information Retrieval analysis. Various music features are used as input to several classifiers, including neural networks. MFCC and tonal features are found to be the best predictors of genre.

\section{Design of EMIR}
\subsection{Outline}
The Ethiopian Music Information Retrieval dataset contains samples of the four main Ethiopian pentatonic scales (Kiñits): Tizita, Bati, Anchihoye, and Ambassel. Spiritual and secular songs based on these scales were collected and each was assigned to its most appropriate scale, by experts on Yared music. As previously noted, music in each scale also has distinctive stylistic characteristics, so Kiñit identification is related to genre classification for other forms of music. Classification accuracy was measured by inter-annotator agreement. Finally, music recordings were labeled and grouped together to form the EMIR dataset.

\begin{table}[ht]
  \centering
        \caption{Number of utterances per speaker in the ASED database.}
%        \resizebox{\columnwidth}{!}{%
\begin{tabular}{llccccc} 
\hline 
Type & Source & \multicolumn{4}{c}{Kiñit (Genre)}  & Total                                    \\ \cline{3-6}
%\multirow{}{}{Id} & \multirow{}{}{Emotion} & \multicolumn{4}{c}{Amharic dialect}  & \multirow{}{}{Number of recordings} &                  &                   \\ \cline{3-6}
                    &                                    & Tizita  & Bati  & Ambassel  & Anchihoye  &                                         \\ \hline
\begin{tabular}[c]{@{}l@{}}Ethiopian Orthodox Tewahedo\\ chants accompanied by traditional\\ instruments\end{tabular}                  & YouTube, DireTube                            & 10      & 6    & 12    & 5      & 33                                     \\
\begin{tabular}[c]{@{}l@{}}Songs performed in Azmari houses\\ accompanied by traditional\\ instruments\end{tabular}                    & Recorded by musicologists           & 7     & 8    & 11    & 4       & 30                                     \\

Azmari Songs                 & YouTube, DireTube                                & 5    & 2     & 6      & 3       & 16                                     \\
Secular Music                   & YouTube, DireTube                              & 140      & 128    & 118    & 135     & 521                                    \\ \hline  
Total                   &                               & 162      & 144    & 147    & 147     & 600                                    \\
 \hline                               
\end{tabular}
%}
\label{four main Ethiopian pentatonic scales}
\end{table}

\subsection{Recordings}

There are three types of recording in EMIR: 
Firstly, Ethiopian Orthodox Tewahedo chants, which form part of a religious tradition dating back to the time of Saint Yared, secondly traditional Ethiopian Azmaris (songs), and thirdly modern secular Ethiopian music. 

The Orthodox chants were collected from online sources such as YouTube and DireTube. Some Azmaris were specially recorded in Addis Ababa by an ethnomusicologist specialising in Azmari houses; these are traditional venues where Azmaris are studied and performed. Firstly, five typical Azmari houses were selected for the study. Secondly, these were visited on multiple occasions. Each time, a singer was asked whether they would record an Azmari of their choice which was in a specified Kiñit. If the singer knew an Azmari in that Kiñit and they agreed to the recording, it went ahead. Otherwise, another singer was asked. 
In this way, over several visits to each house, a collection of Azmaris in the different Kiñits was built up.

The Azmaris were recorded with an AKG Pro P4 Dynamic microphone, at a distance of 25 cm from the singer's mouth. The audio file was saved at a 16 kHz sampling rate and 16 bits, resulting in a mono .wav file.  We used the Audacity audio editing software \cite{audacity:xxx} to reduce the background noise of the music signal. 

Further Azmaris were collected from online sources such as YouTube etc.
Finally, the secular music was collected from online sources. The breakdown of recordings can be seen in Table \ref{four main Ethiopian pentatonic scales}. In all cases, music clips in EMIR are limited to 30 seconds length in order to protect the copyright of the originals. 
\subsection{Judgements}
Five judges participated. Two of them were Ethiopian postgraduate students of Computer Science at Xidian University %(\zh{西安电子科技大学}), China.
Three further judges were from the Yared music school in Addis Ababa, Ethiopia. All five were experts on Yared music. Judges were responsible for the quality control of the dataset.

Each Judge listened independently to all the recordings. For each one, they either assigned it to one of the four Kiñits, or rejected it as not clearly falling into any one of them. If three or more judges assigned a recording to the same category, it was accepted for EMIR. Otherwise it was rejected.

Since we had five judges, the Fleiss kappa \cite{randolph2005free} coefficient was used to calculate the pairing agreement between participants:
   
 \begin{equation} \label{eq:1}
\kappa=\frac{\overline{p}_{0}-\overline{p}_{e}}{1-\overline{p}_{e}} 
\end{equation}

The factor ${1-\bar{p}_{e}}$ gives the degree of agreement that is attainable above chance, and ${\bar{p}_{0}-\bar{p}_{e}}$ gives the degree of agreement actually achieved above chance: $k=1$, if all the raters are in complete agreement. Evaluation of the inter-rater agreement for our dataset in terms of Fleiss kappa is 0.85. This value shows a high agreement level among our five raters.
\subsection{Files and Labeling}
The tracks are all 16 KHz Mono 16-bit audio files in .wav format. 
Each file was labeled in the form `Bati1.Wav'. The first part of the name indicates the Kiñit (Tizita, Bati, Ambassel or Anchihoye); the second part indicates the number of the recording within that Kiñit (1, 2, 3...). 
Subsequently, recordings were stored in four different folders in the dataset. 

We aimed to collect 1,000 recordings. However, because of the above selection process, many of these were rejected, resulting in 600:
162 Tizita, 144 Bati, 147 Ambassel and 147 Anchihoye (Table \ref{four main Ethiopian pentatonic scales}).

EMIR was split into training, validation and testing sets randomly. The training set contains 70\% of the whole dataset, the validation set contains 10\% and the testing set 20\%.
The EMIR dataset is freely available for research purposes\footnote{\url{https://github.com/Ethio2021/EMIR_Dataset_V1}}.
\section{FEATURE EXTRACTION}\label{FEATURE}

The main goal of the feature extraction step is to compute a sequence of feature vectors, providing a compact representation of the given input signal.
In music classification, feature extraction requires much attention because classification performance depends heavily on it.

In previous studies on music genre classification, four feature types have been employed: FilterBanks  \cite{lim2011music, munoz2017nonnegative}, Mel-spectrograms (MelSpec) \cite{chillara2019music,dhall2021music,tang2020combining,ghildiyal2020music}, Chroma \cite{zhang2014verification, singh2021deep, pulipati2021music}, and MFCC \cite{kizrak2014classification,thiruvengatanadhan2018music,mandal2020automatic,sharma2021classification}. These four are all used in our experiments to determine which can perform better for EMIR data.

\subsection{FilterBanks}
A Mel FilterBank is a triangular filter bank that works similarly to the human ear's perception of sound; thus it is more discriminative at lower frequencies and less discriminative at higher frequencies. Mel FilterBanks are used to provide a better resolution at low frequencies and less resolution at high frequencies \cite{fayek2016}.
\subsection{Mel-Spectrograms (MelSpec)}
The signal is separated into frames and a Fast Fourier transform (FFT) is calculated for each frame. A Mel-scale is then created, where the entire frequency spectrum is separated into evenly spaced bands. A spectrogram is then created where, for each frame, the signal magnitude is decomposed into its components, corresponding to the frequencies in the Mel-scale.

\subsection {Chroma} 
The chroma feature is widely used in Music Information Retrieval \cite{bartsch2001catch}. It is made in accordance with the twelve-tone Equal Temperament. Because notes exactly one octave apart are perceived as very similar in music, knowing the distribution of the Chroma even without the absolute frequency (i.e. the original octave) provides important musical information about the audio, and may even show perceived musical similarities not visible in the original spectra. Chroma features are usually represented as a 12-dimensional vector $v=[V(1), V(2), V(3),...V(12)]$; each element of the vector is connected with one element of the set {C, C\#, D, D\#, E, F, F\#, G, G\#, A, A\#, B}, representing the local energy distribution of the audio signal at semitones represented by the 12 pitch names. 
\subsection {Mel-Frequency Cepstral Coefficients (MFCC)} 
Mel-frequency cepstral coefficients (MFCC) are widely employed to extract features from sound in various applications such as speech recognition \cite{glittas2021low},
Music Emotion Recognition \cite{dutta2021music}, and music genre classification  \cite{kizrak2014classification,thiruvengatanadhan2018music,mandal2020automatic,sharma2021classification,1416349}.
MFCC is designed using knowledge of the human auditory system, and is a common method for feature extraction in speech recognition. However, it can also be used with music because it simulates the function of the human ear \cite{li2011genre}. Extracting features with MFCC involves splitting the signal into short frames and then, for each frame, calculating the periodogram estimate of the power spectrum. The Mel FilterBank is applied to the power spectra, to collect the energy for each filter. The log energies of all FilterBanks are calculated, and hence the Discrete Cosine Transform (DCT) of log FilterBank energies is determined. Finally, DCT coefficients 13-20 are saved, with the rest being removed.
\subsection{Extraction}
Initially, each music clip in the data set is divided into specific lengths of time window with 50\% overlap, i.e. 1s, 3s or 5s.
A feature is then created for each window, resulting in a feature vector.
A feature vector can contain FilterBank, Chroma, MelSpec or MFCC features.
The feature vector for each music clip is used within the proposed EKM model for classification. 
MFCCs are extracted using 40 Mel-bands and 13 DCT coefficients.
        
\section{NETWORK ARCHITECTURES and SETUP} \label{architectures}

\subsection{Existing Classification Architectures}
	As discussed earlier, most previous studies employ CNN-based models such as AlexNet, VGG or ResNet, or LSTMs for sound classification \cite{simonyan2020very}. The following is a short overview of these models.
		\begin{itemize}	
			\item \textbf{AlexNet} \cite{krizhevsky2012imagenet} was the first CNN-based model to be used in the ImageNet competition, in 2012.  AlexNet's success launched a revolution, enabling numerous complex tasks to be solved with better performance. It has been widely used for music classification \cite{feng2017music,yang2020parallel,dhall2021music}
			
			\item \textbf{VGG} \cite{simonyan2014very, xie2019investigation} networks appeared in 2014, developed by Oxford Robotics Institute. They were the first to employ considerably smaller 3 × 3 filters in each convolutional layer, furthermore combining them as a convolution sequence. MIR applications include Shi et al. \cite{shi2019music} and Das et al. \cite{das2019double}.
			
			\item \textbf{ResNet} \cite{he2016deep} was launched in late 2015. This was the first time that networks having more than one hundred layers were trained. Subsequently it has been applied to music classification \cite{das2019double,li2021evaluation}.
			\item \textbf{LSTM} is a form of recurrent network which has been successfully used for music classification \cite{tang2018music,deepak2020music,yi2021music}.
		\end{itemize}

\begin{figure}
			\centering
			\includegraphics[width=100mm,scale=1.5]{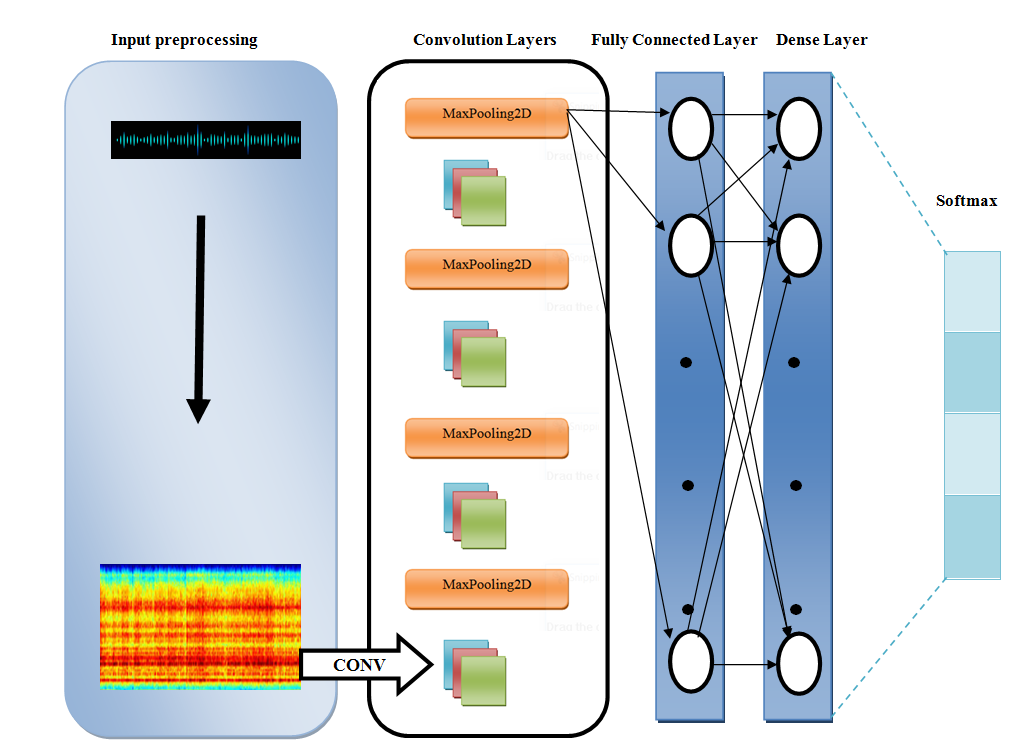}
			\caption{ EKM architecture used in experiments.}
			%\Description{ Training curves of the Val-accuracy of the models LSTM, Alex-Net, ResNet50 and VGG on ASED.}
			\label{vgg-style-network}
		\end{figure}
\subsection{Proposed EKM Architecture}
As we have mentioned, VGG is one of the earliest CNN models used for signal processing. It is well known that the early CNN layers capture the general features of sounds such as wavelength, amplitude, etc., and later layers capture more specific features such as the spectrum and the cepstral coefficients of waves. This makes a VGG-style model suitable for the MIR task.

VGG16 consists of 13 convolution layers with 3x3 kernels, 5 MaxPooling layers with pool size 2x2 filters, 2 fully connected layers, and finally a softmax layer.
We therefore developed the Ethio Kiñits Model (EKM) for this work, based on VGG. After some experimentation, we found that a four-layer architecture gave the best performance.
EKM thus consists of 4 convolution layers with sizes 32, 64, 128, and 256, respectively, with kernels 3x3 for the first three convolution layers and  2x2 for the last convolution layer. There are also 4 MaxPooling layers with pool size 3x3 for the first  MaxPooling layer and 1x1 for the remaining layers. Finally, there is a fully connected layer and a softmax layer.
The proposed model is shown in Figure \ref{vgg-style-network}. AlexNet, ResNet50, VGG16 and LSTM were also used for comparison.

\subsection{Experimental Setup}

The standard code for AlexNet, ResNet50, VGG16 and LSTM was downloaded and used for the experiments. For EKM, the network configuration was altered (Figure \ref{vgg-style-network}). For the other models, the standard network configuration and parameters were used.

In all experiments, the librosa v0.7.2 library \cite{mcfee2015librosa} was used to extract FilterBank, MelSpec, Chroma and MFCC features.

We used the Keras deep learning library (version 2.0), with Tensorflow 1.6.0 backend, to build the classification models. The models were trained using a machine with an NVIDIA GeForce GTX 1050. The Adam optimization algorithm was used, with categorical cross-entropy as the loss function; training stopped after 250 epochs, and the batch size was set to 32.		
\section{EXPERIMENTS} 
\subsection{EXPERIMENT 1: CHOICE OF FEATURES}\label{experiment1}
The aim of the first experiment was to choose the most efficient technique to use for extracting features from the proposed dataset. As we have discussed in Section \ref{FEATURE}, FilterBank, MelSpec, Chroma and MFCC are four forms of feature that are widely used within MIR systems. We therefore wished to determine which of these was most suitable for Kiñit classification on EMIR.

VGG CNN-based models have performed very well for other music. Therefore, the proposed EKM architecture is based on VGG, as discussed earlier. Training and testing were performed with EMIR data using 3s samples. We extracted features using four methods, FilterBank with 40 bands, MelSpec with 128 bands, Chroma with 12 bands and MFCC with 40 bands. First, the model was trained and evaluated using just Filterbank features. Training and evaluation were then repeated using just MelSpec, Chroma and MFCC features.

Data was split 70\% train 10\% validation and 20\% test. The model was trained five times and the average result was reported.

\begin{figure}
\centering
\begin{subfigure}{.24\textwidth}
    \centering
    \includegraphics[width=1.1\linewidth]{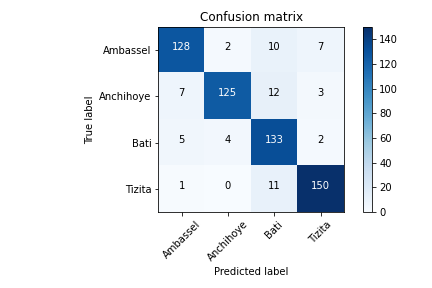}  
    \caption{FilterBank}
    \label{SUBFIGURE LABEL 1}
\end{subfigure}
\begin{subfigure}{.24\textwidth}
    \centering
    \includegraphics[width=1.1\linewidth]{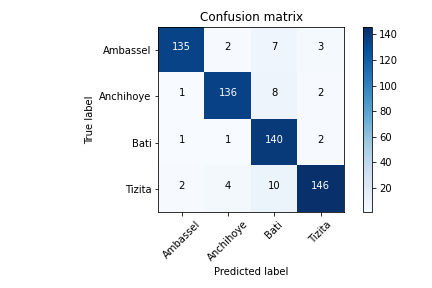}  
    \caption{MelSpec}
    \label{SUBFIGURE LABEL 2}
\end{subfigure}
\begin{subfigure}{.24\textwidth}
    \centering
    \includegraphics[width=1.1\linewidth]{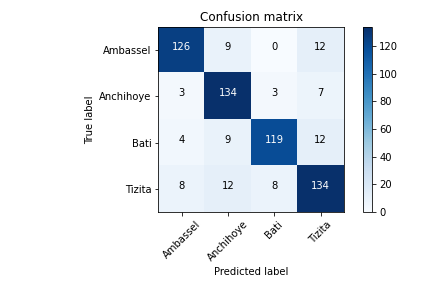}  
    \caption{Chroma}
    \label{SUBFIGURE LABEL 3}
\end{subfigure}
\begin{subfigure}{.24\textwidth}
    \centering
    \includegraphics[width=1.1\linewidth]{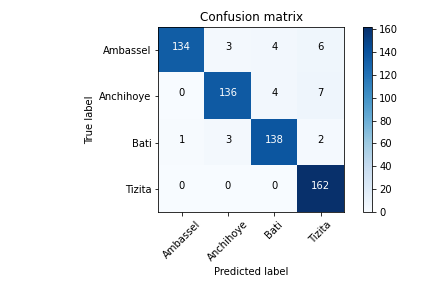}  
    \caption{MFCC}
    \label{SUBFIGURE LABEL 4}
\end{subfigure}
\caption{Experiment 1: EKM confusion matrices using FilterBank, MelSpec, Chroma and MFCC on EMIR.}
\label{fig:foobar}
\end{figure}

The results are presented in Table \ref{recog-accuracy}. As can be seen, MFCC outperforms the other three methods with classification accuracy of 95.00\% as compared with 92.83\% for MelSpec, 89.33\% for FilterBank, and 85.50\% for Chroma. Therefore, we used MFCC processed data for subsequent experiments.
After inspecting the overall results, we decided to interpret the performance on a genre level by comparing the genre classification confusion matrices arising from the model when trained with the four different types of feature (Figure \ref{fig:foobar}). The vertical axis represents the ground truth and the horizontal axis represents the prediction. The diagonal lines in all four matrices show that predictions in general reflect the ground truth.
It is also clear that the network performs better with some genres than others. Figures \ref{fig:foobar} (a), (b) and \ref{fig:foobar} (d) (FilterBank, MelSpec and MFCC) show that the Bati, Tizita and Anchihoye scale was always easily identifiable, while Ambassel was hard to identify. 
Looking at the confusion matrices in more detail,
in Figure \ref{fig:foobar} (a), the FilterBank EKM model incorrectly classifies 10 Ambassel cases as Bati and 7 as Tizita. In consequence, the number of correct predictions for the Ambassel class is reduced to 128. FilterBank also shows less gains in predicting 12 Anchihoye as Bati as compared to MFCC, where 4 Anchihoye are predicted as Bati. This results in the correct prediction of Anchihoye being 125, relative to 136 for MFCC. This outcome appears to be conceivable because MFCC can benefit from the difference between the genre distributions of Bati and Tizita expressions. It is striking that the FilterBank EKM model incorrectly predicts 11 of the Tizita class as Bati, 7 of the Ambassel as Anchihoye, and 5 of the  Ambassel as Bati.
The MelSpec model, Figure \ref{fig:foobar} (b), shows less prediction gains in predicting 10 Tizita as Bati. In consequence, 146 Tizita are correctly classified as compared to 162 for MFCC. The model incorrectly classifies 8 Bati cases as Anchihoye and 7 Bati as Ambassel.

In Figure \ref{fig:foobar} (c), the Chroma model incorrectly classifies 8 Tizita cases as Bati and Ambassel. Moreover, it predicts 12 Ambassel as Tizita. Thus, only 126 Ambassel are correctly classified, as compared to 136 for MFCC. It is striking that the model incorrectly predicts 12 Bati as Tizita and 12 Tizita as Anchihoye.

Compared to the other three features, the MFCC model in Figure \ref{fig:foobar} (d) shows significant gains when predicting the Anchihoye class (136 correct, vs. 125 for FilterBank) and the Tizita class (162 correct, vs. 146 for MelSpec). MFCC never predicts Ambassel as Anchihoye, compared to 7 times for FilterBank. Also, it predicts 0 Tizita as Bati, compared to 10 for MelSpec. The frequency of incorrect cases Ambassel-to-Tizita, Anchihoye-to-Tizita and Bati-to-Tizita relative to the Chroma decreased from 8 to 0, from 12 to 0 and from 8 to 0 respectively. The values on the diagonal axis indicate that the accuracies of the correctly predicted class have increased.

Generally, classification performance was somewhat inconsistent across genres. While Bati, Tizita, and  Anchihoye music are comparatively distinct, Ambassel are often ambiguous. This discovery is consistent with human performance, as it is more difficult for humans to identify some genres than others \cite{jothilakshmi2012automatic}.

This also suggests that the choice of Kiñit could greatly affect the difficulty of a classification task. A dataset consisting of Tizita tracks would be significantly easier to classify, while Anchihoye would be more difficult.
\begin{table}[ht]
\centering
\caption{Experiment 1: Recognition Accuracies of VGG Networks on EMIR Using FilterBank, MelSpec, MFCC and  Chroma Features with 3s samples.}
\begin{tabular}{cccccc}
 \hline
\multicolumn{1}{c}{Dataset} & \multicolumn{4}{c}{Features}                      & Approach             \\ \hline
\multirow{2}{*}{EMIR}       & FilterBank & MelSpec & Chroma  & MFCC & \multirow{2}{*}{EKM} \\\cline{2-5}
                            & 89.33      & 92.83           & 85.50 &  95.00 &       \\ \hline               
\end{tabular}
 \label{recog-accuracy}
\end{table}
%+++++++++++++++++

 %         \FloatBarrier 
%\begin{figure}[ht]
%			\begin{subfigure}{.4\textwidth}
%				\centering
				% include first image
%				\includegraphics[width=.8\linewidth]{fig/600/fbank.png}  
%				\caption{FilterBank}
				%\label{fig:sub-first}
%			\end{subfigure}
%				\begin{subfigure}{.4\textwidth}
%				\centering
				% include first image
%				\includegraphics[width=.8\linewidth]{fig/600/Mel.png}  
%				\caption{MelSpec}
				%\label{fig:sub-first}
%			\end{subfigure}
%				\begin{subfigure}{.4\textwidth}
%				\centering
				% include first image
%				\includegraphics[width=.8\linewidth]{fig/600/Mel.png}  
%				\caption{MelSpec}
				%\label{fig:sub-first}
%			\end{subfigure}
%				\begin{subfigure}{.4\textwidth}
				%\centering
%				% include first image
%				\includegraphics[width=.8\linewidth]{fig/600/Mel.png}  
%				\caption{MelSpec}
				%\label{fig:sub-first}
%			\end{subfigure}
			
%			\caption{VGGb confusion matrices using Mel-Spectogram and MFCC features on  text-dependent ASED.}
%			\label{fig:foobar}
%	 	\end{figure}    
 %\FloatBarrier 
%+++++++++++++++++++++++++++++++++++++++++++++++++
 \FloatBarrier 
 \FloatBarrier

%+++++++++++++++++++++++++++++++++++++++++++++++++

%\begin{figure}
 %   \centering
%    \subfigure[FilterBank]{\includegraphics[width=0.4\textwidth]{fig/600/fbank.png}} 
%    \subfigure[ MelSpec]{\includegraphics[width=0.4\textwidth]{fig/600/Mel.png}} 
%    \subfigure[Chroma]{\includegraphics[width=0.4\textwidth]{fig/600/chroma.png}}
%    \subfigure[MFCC]{\includegraphics[width=0.4\textwidth]{fig/600/MFCC_1.png}}
%    \caption{Experiment 1: EKM confusion matrices using FilterBank, MelSpec, Chroma and MFCC on EMIR.}
%    \label{fig:foobar}
%\end{figure}

\subsection{EXPERIMENT 2: CHOICE OF SAMPLE LENGTH} \label{experiment2}
From a human perspective, it usually takes only a few seconds to determine the genre of an audio excerpt. Therefore, short samples were used, having a length of 1s, 3s or 5s. The aim of Experiment 2 was to choose the optimal sample length for the Kiñit classification of EMIR data. 

Three variants of the EKM model were created, one using 1s samples for all clips, one using 3s and one using 5s. Each model was trained five times using a 70\%/10\%/20\% train/validation/test split and the average results were computed.

Results are presented in Table \ref{comparison-df-Sample Lengeth}. EKM had the highest accuracy on sample length 3s (95.00\%), sample length 1s being close behind (94.44\%). Sample length 5s was the worst (90.28\%).

\begin{figure}
        	\centering
        	\includegraphics[width=65mm,scale=1.5]{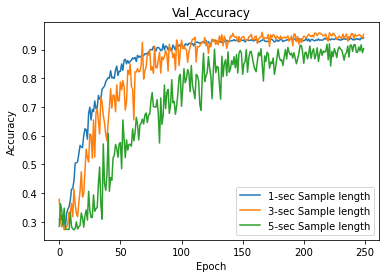}
        	\caption{Experiment 2: Convergence curve in 250-epoch training.}
        	%\Description{Convergence Curve in 500-epoch %training.}
        	\label{Experment2}
        \end{figure}

Figure \ref{Experment2} shows the Val-accuracy curve for the three models, having sample length 1s, 3s and 5s. The models are trained for 250 epochs. The curves show that after the 150th epoch, the Val-accuracy starts stabilizing. The curve for three seconds looks like a better fit, while that for five seconds shows more noisy movements than the other sample lengths.

\begin{table}[ht]
        \centering
        \caption{Experiment 2: EKM model accuracy using 1s, 3s and 5s sample lengths, MFCC features and EMIR data.}
\begin{tabular}{cccccccccc}
\hline
\multicolumn{3}{l}{Dataset}                & Approach              & Features                & Number of genres   &  Sample Lengeth       & \multicolumn{3}{l}{Accuracy} \\\hline
\multicolumn{3}{l}{\multirow{3}{*}{EMIR}} & \multirow{3}{*}{EKM} & \multirow{3}{*}{ MFCC} & \multirow{3}{*}{4} & One second & \multicolumn{3}{l}{94.44\%}     \\\cline{7-8}
\multicolumn{3}{l}{}                       &                       &                                &                    & Three seconds            & \multicolumn{3}{l}{95.00\%}     \\\cline{7-8}
\multicolumn{3}{l}{}                       &                       &                                &                    & Five seconds            & \multicolumn{3}{l}{90.28\%}  \\\hline  
\end{tabular}
\label{comparison-df-Sample Lengeth}
\end{table}

\subsection{EXPERIMENT 3: COMPARISON OF GENRE CLASSIFICATION MODELS}\label{experiment3}
The aim was to compare four established models (Section \ref{architectures}) with the proposed EKM model, when applied to Kiñit classification.  Recall that the four models are AlexNet, ResNet50, VGG16 and LSTM. Once again, MFCC features were used. The network configuration for EKM was the same as in the previous Experiment (Figure \ref{vgg-style-network}). For the other models, the standard configuration and settings were used.

\begin{table}
        \centering
        \caption{Experiment 3: Comparison of EKM with other CNN and LSTM models, all applied to the EMIR dataset.}
           	 \begin{tabular}{ccccc}
           		 \hline  
           	No.	& Models& Training completed in time & Accuracy  \\\hline 
           	1	&  LSTM  &  00:08:46 &87.50\%\\
            2	&  AlexNet  & 01:09:41  &89.83\% \\
            3	&  ResNet50  & 01:37:04  &90.50\% \\
            4	&  VGG16  & 01:34:09  &93.00\% \\
           	5	&  EKM  & 00:09:17  &95.00\% \\
           		\hline 
           	\end{tabular}
           	\label{comparison-vgg-cnn-rnn}
           \end{table}   
Results are presented in Table \ref{comparison-vgg-cnn-rnn}. EKM had the highest accuracy (95.00\%), VGG16 being close behind (93.00\%).
In addition, EKM was also much faster than VGG16 (00:09:17 vs. 01:34:09), showing that it is more efficient and hence more suitable for applying to MIR datasets. 

\section{CONCLUSIONS}
   In this paper, we first collected what we believe to be the very first MIR dataset for the Ethiopian music, working with four main pentatonic Kiñits (scales), Tizita, Bati, Ambassel and Anchihoye. We then conducted three experiments. The first experiment was to determine whether Filterbank, MelSpec, Chroma, or MFCC features were most suitable for genre classification in Ethiopian music. When used as the input to the EKM model, MFCC resulted in superior performance relative to Filterbank, MelSpec and Chroma (95.00\%, 89.33\%, 92.83\% and 85.50\%, respectively) suggesting that MFCC features are more suitable for Ethiopian music.
   In the second experiment, after testing several sample lengths with EKM and MFCC features, we found the optimal length to be 3s.
   In the third experiment, working with MFCC features and the EMIR data, we compared the performance of five different models, AlexNet, ResNet50, VGG16, LSTM, and EKM. EKM was found to have the highest accuracy (95.00\%) as well as the second shortest training time (00:09:17).
Future work on EMIR includes enlarging the scale of the database using new elicitation techniques, and studying further the effect of different genres on classification performance.

\section*{Acknowledgments}
    This work was supported by the National Key Research and Development Program of China under grant 2020YFC1523300.

\bibliographystyle{abbrv}
%\bibliography{references} 

\end{document}